\documentclass[twocolumn,showpacs,preprintnumbers,amsmath,amssymb]{revtex4}
\usepackage{graphicx}
\usepackage{subfigure}
\usepackage{amsbsy}
\begin{document}

\bibliographystyle{apsrev}

\title{Field autocorrelations in electromagnetically induced transparency:
  Effects of a squeezed probe field}
\author{P. Barberis-Blostein  }
\affiliation{Centro de Ciencias F\'\i sicas, Universidad Nacional Autonoma de
M\'exico, Av. Universidad s/n, Colonia Chamilpa, Cuernavaca, Morelos}
\date{today}
\begin{abstract}
The interaction of a quantized field with three-level atoms in $\Lambda$
configuration inside a two mode cavity is analyzed. We calculate the
stationary quadrature
noise spectrum of the field outside the cavity in
the case where the input probe field is in a squeezed state and the atoms show
electromagnetically induced transparency (EIT). If the
Rabi frequencies of both dipole transitions of the atoms are different from
zero, we show that the output probe field have four maxima of squeezing absorption. We show that in
some cases two of these frequencies can be very close
to the transition frequency of the atom, in a region where the mean value of
the field entering the cavity is hardly altered.
Furthermore, part of the absorbed squeezing of the probe field is transfered
to the pump field. For some conditions this transfer of squeezing can be complete.
\end{abstract}
\pacs{42.50.Gy,42.50.Lc,42.50.Ar,42.50.Pq}  
\maketitle
\vskip2pc

\section{Introduction}
\begin{figure}[t]
\includegraphics[width=3in]{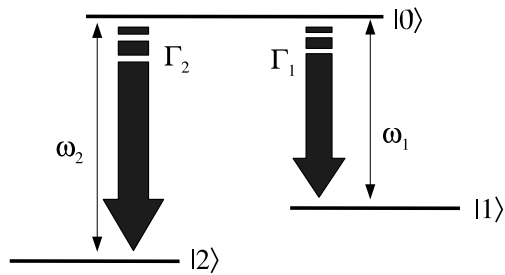}
\caption{\label{fig:atomo} Atom in $\Lambda$ configuration, $\Gamma_1$ and
  $\Gamma_2$ represent radiative decaying constants.}
\end{figure}

Electromagnetically induced transparency (EIT) \cite{rv:harris} is a technique
that can be used to eliminate the fluorescence from an atom
illuminated with light whose frequency is equal to a particular atomic
transition. This phenomenon has been observed in systems of three-level atoms
in $\Lambda$ configuration (see Fig.~\ref{fig:atomo}) \cite{rv:marangos}. In this configuration a mode of the field, called the pump field, interacts resonantly
with one dipole transition, while another mode, the probe field, interacting with the second dipole transition, is tested for transparency. The
maximum absorption of the probe field by the medium depends on the Rabi
frequencies associated with each atomic optical transition. The maximum occurs
for a detuning from resonance which increases monotonically with the Rabi frequencies. EIT has many applications such as ultra slow
propagation \cite{rv:hau} and storage of light \cite{rv:liu}, to name just a
few. 

In general, the mean value of the electromagnetic field
after interacting with the medium is measured. We may wonder if
the medium is also transparent to the field fluctuations of
the initial incoming probe field,
particularly if these fluctuations are due to an initial state of the field which
does not have a classical analog, such as squeezed states. In order to
further investigate this question it is necessary to treat the field quantum mechanically. Under the assumption
that the pump field is classical and with Rabi frequency much larger than the
associated Rabi frequency of the probe field treated quantum mechanically,
Fleischhauer et.\,al.\,\cite{rv:memoria2} showed that the
medium is transparent to the full quantum state. Furthermore they showed that they
can transfer the state of the probe field to the atoms and then back to the
field again. These results
were partially confirmed by the experimental work of Akamatsu
et.\,al.\,\cite{rv:kozuma}, where the transparency of the atoms was measured
considering an
initially squeezed vacuum state as a probe field.
Recent experimental \cite{rv:carlosruido} \cite{rv:lezama} \cite{rv:scullyeitco} 
studies have shown how the noise of the probe field is influenced by the medium of
$\Lambda$ systems in a parameter region where the atoms show
electromagnetically induced transparency. Also theoretical work where the pump and probe
fields are treated quantum mechanically 
\cite{rv:pablonicim} \cite{rv:dantan3} \cite{rv:dantan4} \cite{rv:dantan5} has recently been done. It turns out that
when both modes, pump and probe, are in a coherent state, the medium is transparent for the initial
fluctuations of the coherent state, except for expected fluorescence due to
small absorption from the media when they are not in the totally ideal EIT
situation \cite{rv:pablonicim}.  Moreover, the effects of the field noise,
particularly quantum noise and its influence on the atoms' state has been investigated. Dantan et.\,al.\,\cite{rv:dantan3}
\cite{rv:dantan4} have shown that an initially squeezed vacuum as a probe
field can be transferred to the atomic ensemble. The
situation where the Rabi frequency of the pump and probe field are comparable
and the probe field is in a general squeezed state still lacks a complete
theoretical treatment.

In this paper we study the statistical properties of the electromagnetic
quadrature functions of the stationary output field from a
two mode cavity filled with three-level atoms in $\Lambda$ configuration. In
particular we investigate the case when the two modes of the cavity field are in resonance with the dipole transitions of
the three-level atoms and the corresponding interaction constants and
decay rates are equal. For this case, when the mean value of the pump field is not zero the atoms show EIT and the average
values of the field do not change during the interaction. We focus on the case
where the input probe field does not have a classical analog, namely when it
is in a
squeezed state. 
The most interesting feature of the results for the quadrature noise spectrum
is that, for a given frequency,
close to the transition frequency of the atoms associated to the probe
field, we have a maximum of absorption of the initial squeezing of the probe
field. Remarkably, some of this initial squeezing of the probe field can be transferred
to the pump field. This transfer is maximized when both Rabi frequencies are
equal. The spectrum frequency where this transfer happens is
very close to the resonance frequency of the probe field with the atom.
Our results imply that the initial quantum properties of the fields are
modified after interacting with the medium showing EIT.

This paper is organized as follows: in section~\ref{sc:equation}
we give a brief review on how the system equations are obtained and solved. In
section~\ref{sc:analysis} we present the main new results, which consist of
analytical expressions for the spectrum of the quadrature noise of the field after
interacting with the medium when the Rabi frequency associated with each field
is different from zero. Finally we present our conclusions in section~\ref{sc:conclusions}.

\section{Cavity output field equations}
\label{sc:equation}
Consider the case of $N$ three-level atoms in $\Lambda$ configuration
inside a cavity that sustains two modes of the electromagnetic field. The
frequencies of these modes are resonant with the transitions between levels
$|1\rangle$ and $|0\rangle$ and
$|2\rangle$ and $|0\rangle$ of the atom, respectively. The atoms are supposed
to occupy a volume of small dimension as compared with the wavelength of the
cavity modes. We point out that this is a difficult task to realize experimentally. The annihilation field operators for the modes inside the cavity are denoted
by $\hat a_1$ and $\hat a_2$.  We will work in the
interaction picture, where the Hamiltonian of the cavity-atom interaction is
given by
\begin{eqnarray}
H_{\textrm{int}}&=&\sum_{j=1}^N\hbar (g_1\hat{\sigma}_{01}^j\hat a_{1}
+g_2\hat{ \sigma}_{02}^j\hat a_{2}+ \textrm{c.c.}
\,)\nonumber\\
&=&\hbar g_1\hat{\Sigma}_{01}\hat a_{1}
+\hbar g_2\hat{ \Sigma}_{02}\hat a_{2}+ \textrm{c.c.}
\, ,
\label{eq:hamilcav}
\end{eqnarray}
where $\hat{\Sigma}_{ij}=\sum_{k=1}^N\hat{\sigma}_{ij}^k$ are the collective
operators that represent the sum of individual atomic operators,
$\sigma_{ij}^k=|i\rangle\langle j|^k$ associated with the $k^{\rm th}$ atom. We use input-output theory to relate the inside
field with the outside field \cite{lb:wallsmilburn}. Each intra-cavity mode interacts with its
own collection of modes in the outside field. This means that either they have different
polarizations or their difference in frequency is large. The intracavity operators
satisfy the equations
\begin{eqnarray}\label{eq:cavidad}
\frac{d}{dt}\hat a_1(t) & = & -i g_1 \hat \Sigma_{10}(t)-\frac{\gamma_1}{2}
\hat a_1(t)+\sqrt{\gamma_1}\,\hat a_{1{\rm in}}(t)\, ,\nonumber\\
\frac{d}{dt}\hat a_2(t) & = & -i g_2 \hat \Sigma_{20}(t)-\frac{\gamma_2}{2}
\hat a_2(t)+\sqrt{\gamma_2}\,\hat a_{2{\rm in}}(t)\, ,
\end{eqnarray} 
where $\gamma_i$ is the decay rate of cavity mode $i=1,2$.
The operators $\hat a_{i{\rm in}}(t)=-1/\sqrt{2\pi} \int_{-\infty}^{\infty}d\omega
e^{-i\omega(t-t_0)}\hat b_i(t_0,\omega)$ represent the
field entering the cavity. The operator
$\hat b_i(t_0,\omega)$ represents the outside mode associated with cavity mode
$i$ at the initial time $t_0$ and frequency
$\omega$.
We will call the outside field associated with the modes labeled by index $i=1$
($i=2$) the pump (probe) field.
In the interaction picture
$\omega$ represents the detuning from the cavity frequency. The outcoming field is given by $\hat a_{i{\rm out}}(t)=1/\sqrt{2\pi} \int_{-\infty}^{\infty}d\omega
e^{-i\omega(t-t_1)}\hat b_i(t_1,\omega)$. The operators $\hat b_i(t_1,\omega)$
represents the outside mode associated with cavity mode
$i$ at time $t_1>t_0$ and frequency
$\omega$ \cite{lb:wallsmilburn}. 
The incoming and outcoming
fields are related by

\begin{eqnarray}\label{eq:io}
\hat a_{1{\rm in}}(t)+\hat a_{1{\rm out}}(t)&=&\sqrt{\gamma_1}\hat a_1(t)\, ,\nonumber\\
\hat a_{2{\rm in}}(t)+\hat a_{2{\rm out}}(t)&=&\sqrt{\gamma_2}\hat a_2(t)\, .
\end{eqnarray}

The Langevin equations for the atomic operators are obtained using the
interaction Hamiltonian Eq.~(\ref{eq:hamilcav}). Taking
into account the interaction of the atom with modes other than the cavity in the
usual way \cite{lb:orszag}, we obtain 
\begin{subequations}
\label{eq:sistemacav}
\begin{eqnarray}
\frac{d}{dt}\hat W_1 &=&\frac{1}{3}(-2 \Gamma_1-\Gamma_2)(1+\hat W_{1}+\hat W_2)-2 i g_1\,\hat\Sigma_{01}\hat a_1 
\nonumber\\&&+2 i g_1\,\hat\Sigma_{10}\hat a_1 -i g_2\,\hat\Sigma_{02}\hat a_2 
\nonumber\\&&+i g_2\,\hat\Sigma_{20}\hat a_2  +\hat
F_{W_1}\, , 
\\
\frac{d}{dt}\hat W_2 &=&\frac{1}{3}(-\Gamma_1-2 \Gamma_2)(1+\hat W_{1}+\hat W_2) -i g_1\,\hat \Sigma_{01}\hat a_1 
\nonumber\\&&+i g_1\,\hat \Sigma_{10}\hat a_1  -2 i g_2\,\hat \Sigma_{02}\hat a_2 
\nonumber\\&&+2 i g_2\,\hat \Sigma_{20}\hat a_2  +\hat
F_{W_2}\, ,
\\
\frac{d}{dt}\hat \Sigma_{10}&=&(-\frac{\Gamma_1+\Gamma_2}{2})\hat
\Sigma_{10}+i g_1\,\hat W_1\hat a_1  -i g_2\,\hat \Sigma_{12}\hat a_2 
+\hat F_{10}\, ,\nonumber\\&&
\\
\frac{d}{dt}\hat \Sigma_{20}&=&(-\frac{\Gamma_1+\Gamma_2}{2})\hat
\Sigma_{20}+i g_2\,\hat W_2\hat a_2   -i g_1\,\hat \Sigma_{21}\hat a_1 
+\hat F_{20}\, ,\nonumber\\&&
\\
\frac{d}{dt}\hat{\Sigma}_{21}&=&-i g_1\,\hat{\Sigma}_{20}\hat a_1^\dagger+i
g_2\,\hat{\Sigma}_{01}\hat a_2 \, , \nonumber\\
\end{eqnarray}
\end{subequations}
where $\hat W_j=\hat \Sigma_{00}-\hat \Sigma_{jj}$ and $\hat F_x$ are
collective Langevin operators given by the sum of each atom Langevin operator.

The Langevin fluctuation operators $\hat{F}$'s are assumed
to be delta correlated, with zero mean:
\begin{equation}
\langle \hat F_x\rangle=0\label{langevin}\, ,
\end{equation}
\begin{equation}
\langle \hat F_x(t)\hat F_y(t')\rangle=D_{xy}\delta(t-t')\, ,
\label{eq:langevin}
\end{equation}
where   $x$ and $y$ label the fluctuation operators. 

The atom diffusion coefficients, $D_{xy}$, can be obtained using the generalized Einstein
relations \cite{lb:cohen}. The nonzero diffusion coefficients are given by
Eqs.~(\ref{eq:difcoecav}) in 
appendix~\ref{sc:difusion}. 

We will consider the following initial condition for the incoming field. For
frequencies different from the intra-cavity frequencies, for
each mode of frequency $\omega$ associated to the field  $i$, the field outside the cavity at time
$t_0$ is given by $|\psi(t_0,\omega)\rangle_i=\hat S(r_i,\theta_i)|0\rangle$,
where $|0\rangle$ is the vacuum of the mode with frequency $\omega$ of the field
$i$. The squeeze
operator is given by $\hat S(r_i,\theta_i)=e^{1/2[\beta_i^*\,\hat
  b^2_i(t_0,\omega)-\beta_i\,\hat b^{\dagger 2}_i(t_0,\omega)]}$, with
$\beta_i=r_i\exp{i\theta_i}$. 
When the frequency is equal to the intra-cavity frequency, the
initial condition is: $|\psi(t_0,\omega=0)\rangle_i=\hat D(\alpha_i)\hat S(r_i,\theta_i)
|0\rangle$, where $\hat D(\alpha_i)=e^{\alpha_i\,\hat b^{\dagger}_i(t_0,\omega)-\alpha_i^*\,\hat
  b_i(t_0,\omega)}$ is the displacement
operator which, when applied to the vacuum, creates the coherent state
$|\alpha_i\rangle$. That means that the outside modes are initially in a
vacuum, which can be squeezed, except for the resonant modes with the cavity which
can be in a squeezed state but with field mean value different from zero. We
will chose that value in a way such that inside the cavity we have
$\langle \hat a_i \rangle=\alpha_i$.

Defining the field quadrature $\theta$ for the field $i$ and frequency $\omega$
as
\begin{equation}
\hat Y_{i\, \theta}(\omega,t)=\hat b_i(\omega,t)\exp{(i\theta)}+\hat
b^\dagger_i(\omega,t)\exp{(-i\theta)}\, ,
\end{equation}
we have that the $\theta$-quadrature noise operator is given by
\[
\Delta Y_{i\,\theta_i}(\omega,t=t_0)=\langle(\hat Y_{i\,\theta_i}(\omega,t_0)-\langle\hat
Y_{i\,\theta_i}(\omega,t_0)\rangle)^2\rangle=e^{-2 r_i}\, ,
\]
for the given initial conditions.

Writing $\sqrt{\gamma_i}\,\hat a_{i\, {\rm in}}=\sqrt{\gamma_i}\,\langle \hat a_{i\,
 {\rm in}}\rangle+\hat f_{ai}$, one can show that for the initial conditions
 given above, the operators $\hat f^\dagger_{ai}(t)$ satisfy
\begin{eqnarray}
\langle \hat f^\dagger_{ai}(t)\hat f_{aj}(t')\rangle&=& \gamma_i\,\sinh^2{r_i}\,\delta(t-t') \delta_{ij} \, ,\nonumber\\
\langle \hat f_{ai}(t)\hat f^\dagger_{aj}(t')\rangle&=&\gamma_i(\sinh^2{r_i}+1) \,
\delta(t-t') \delta_{ij}\, ,\nonumber\\ 
\langle \hat f^\dagger_{ai}(t)\hat f^\dagger_{aj}(t')\rangle&=& -\gamma_i\cosh{r_i}\sinh{r_i}\exp{(-i\theta_i)}\,\delta(t-t')\delta_{ij} \, ,\nonumber\\ 
\langle \hat f_{ai}(t)\hat f_{aj}(t')\rangle&=&-\gamma_i\cosh{r_i}\sinh{r_i}\exp{(i\theta_i)}\,\delta(t-t')\delta_{ij}\, .\nonumber\\
\label{eq:campoinicial}
\end{eqnarray}

From Eqs.~(\ref{eq:cavidad}) it is easy to see that $\hat f_{ai}$ represents
the Langevin fluctuation operator associated with the mode $i$ inside the cavity.

Eqs.~(\ref{eq:io}) and~(\ref{eq:sistemacav}) are a set of first order nonlinear
operator stochastic differential equations. To solve this system, it is usual to transform these equations into a
system of c-number Ito stochastic differential equations. These new equations
are equivalent to the original ones up to second order in the operators
\cite{rv:davidovich}. Once we have c-number stochastic equations we can use
normal stochastic methods to solve them \cite{lb:gardiner}. Since c-numbers
commute, in order to perform this operation uniquely, we define an order for
the operator, which
we call ``normal'' order. The normal order we choose is
\begin{equation} 
\hat a_{2}^{\dagger },\hat a_{1}^{\dagger },\hat{\Sigma}_{02},\hat{\Sigma}_{01},\hat{%
\Sigma}_{12},\hat{W}_{1},\hat{W}_{2},\hat{\Sigma}_{21},\hat{\Sigma}_{10},%
\hat{\Sigma}_{20},\hat a_{1},\hat a_{2}\, .
\end{equation}

We will use the c-number variables $\Sigma _{ij}$, $W_{1}$, $W_{2}$, $\alpha
_{i}^{\ast }$, $\alpha _{i}$ for the corresponding operators $\hat \Sigma _{ij}$,
$\hat{W}_{1}$, $\hat{W}_{2}$, $\hat a_{i}^{\dagger }$, $\hat a_{i}.$ The stochastic average of a c-number variable is equal to the mean value of the corresponding operator and the stochastic average of the product of two c-number variables corresponds to the mean value of  the normal ordered multiplication of the two corresponding operators. For example $%
\langle\Sigma _{02}(t)\Sigma _{12}(t^{\prime })\rangle_{st}=\langle\Sigma _{12}(t^{\prime })\Sigma
_{02}(t)\rangle_{st}$ is equal to $\langle \hat{\Sigma}_{02}(t)\hat{\Sigma}%
_{12}(t^{\prime })\rangle $. Here $\langle \cdots \rangle_{st}$ means
stochastic mean. Henceforth we will drop the subscript ``st'' in order to simplify the notation.

The new c-number equations for the system look the same as the operator equations except
that we should replace the Langevin fluctuation operators by  modified
Langevin fluctuation forces.
These modified Langevin fluctuation forces still have zero mean and are still delta function 
correlated. However, the diffusion coefficients associated to these new Langevin fluctuation
forces are modified so that  the
operator equations of normal ordered products coincide with the c-number
equations of the corresponding products.
A very clear explanation of  this procedure is given by L. Davidovich \cite{rv:davidovich}.    

These new normal ordered diffusion  coefficients satisfy the symmetry relation
$D_{xy}=D_{yx}$. The non-zero coefficients (and the symmetrical ones) are
given by Eqs.~(\ref{eq:cnumbercoe}) in the appendix~\ref{sc:difusion}.

We are mostly interested in the dynamics of fluctuations
around the steady state.
In order to calculate these dynamics, we express the solutions of the
stochastic variables as the sum of the  steady state value plus
fluctuations. That is,  for any stochastic variable $O,$ we write $O(t)\approx
\langle O\rangle +\delta O(t),$ with $\delta O\ll\langle O\rangle .$
In the system equations, the atomic operators scale as the number of
atoms, $N$, and the fluctuation forces scale as $\sqrt{N}$ \cite{rv:davidovich}.
If the number of atoms inside the cavity is large
enough ($N\gg 1$) the fluctuations are small and we can neglect terms of order higher than one  in $\delta
O.$ 
Neglecting those terms, we obtain to zeroth order, the equations for the mean values,
which are solved for the stationary state in an analogous way to
bistability problems. The corresponding differential
equations for the fluctuations can be written  in  matrix form
\begin{equation}
\frac{d}{dt}\boldsymbol{\delta O}=\boldsymbol{B}\cdot\boldsymbol{\delta O}+%
\boldsymbol{G}\, ,  \label{eq:fluctuation2}
\end{equation}
where the column vectors $\boldsymbol{\delta O}$ and $\boldsymbol{G}$ have components
\begin{eqnarray}
\boldsymbol{\delta O}^T&=&(\delta \alpha_2^*,\delta \alpha_1^*,\delta \Sigma _{02},\delta \Sigma
_{01},\delta \Sigma _{12},\delta w_{1},\delta w_{2},\delta \Sigma
_{21},\delta \Sigma _{10},\nonumber \\& &\delta \Sigma _{20}, \delta \alpha_1,\delta \alpha_2),
  \nonumber\\ 
\boldsymbol{G}^T&=&(f_{\alpha 1}^*,f_{\alpha 2}^*,f_{02},f_{01},f_{12},f_{w_{1}},f_{w_{2}},f_{21},f_{10},\nonumber\\&
&f_{20},f_{\alpha 1},f_{\alpha 2})\nonumber\, .
\end{eqnarray}
The matrix $\boldsymbol{B}$ can be
obtained from the expansion up to first order in $\delta O$ of the system's
dynamical equations in c-number representation.
The stationary spectrum of the correlation of the c-numbers, $S_{ij}(\omega
)=\int_{-\infty}^\infty d\tau\,e^{i\omega\tau}\langle
O_i(t)O_j(t+\tau)\rangle$ can be written as \cite{lb:orszag}
\begin{equation}\label{eq:defmat}
\langle \delta O_{i}(\omega )\delta O_{j}^{\ast }(\omega
^{\prime })\rangle =S_{ij}(\omega )\delta (\omega +\omega ^{\prime
})\, ,
\end{equation}
where $\langle \dots \rangle $ means stochastic average and $O_{i}(\omega )$ $=\frac{1}{\sqrt{2\pi}}\int_{-\infty }^{\infty }O_{i}(t)\exp {(i\omega t)}%
dt$ is the Fourier transform of $O_{i}(t).$ 

Taking the Fourier transform of Eq.~(\ref{eq:fluctuation2}),
multiplying by $\boldsymbol{\delta O}^{\dagger}$ and taking the stochastic
average, we obtain 
\begin{equation}
\boldsymbol{S}(\omega )=(\boldsymbol{B}+i\,\omega \ensuremath{\,\,\mathrm{l}%
\!\!\!1})^{-1}\cdot\boldsymbol{D}\cdot(\boldsymbol{B^\dagger}-i\,\omega %
\ensuremath{\,\,\mathrm{l}\!\!\!1})^{-1}\, ,  \label{eq:spectrum2}
\end{equation}
where $\boldsymbol{D}\,
\delta(\omega+\omega')=\langle\boldsymbol{G}\cdot\boldsymbol{G}^T\rangle$ and
$\boldsymbol{S}(\omega )$ is the matrix defined by Eq.~(\ref{eq:defmat}). The components of the symmetric diffusion matrix $\boldsymbol{D}$ are the c-number diffusion coefficients given in 
Eq.~(\ref{eq:cnumbercoe}).

The quadrature noise of the output field is given by
\begin{equation}
\Delta Y_{i\,\theta\,\rm out}(t)=(Y_{i\,\theta\,\rm out}(t)-\langle Y_{i\,\theta\,\rm out}(t\rangle)^2 \, .
\label{eq:quout}
\end{equation}
where $Y_{i\,\theta \rm out}=\alpha_{i \,\rm out}
\exp{(i\theta)}+\alpha^*_{i \,\rm out}\exp{(-i\theta)}$. The spectrum of Eq.~(\ref{eq:quout})
can be written as
\begin{equation}\label{eq:quadraturenoise}
\Delta Y_{i\,\theta\rm out}(\omega)\delta(\omega+\omega')=\langle\delta
Y_{i\,\theta \,\rm out}(\omega)  \delta Y_{i\,\theta \,\rm
  out}(\omega')\rangle\delta(\omega+\omega')\, ,
\end{equation}
where $\delta Y_{i\,\theta \rm out}=\delta \alpha_{i \,\rm out}
\exp{(i\theta)}+\delta\alpha^*_{i \,\rm out}\exp{(-i\theta)}$. The calculation
of the latter is our main objective.
Using the c-number equivalents of Eqs.~(\ref{eq:cavidad})  and
Eqs.~(\ref{eq:io}), we have
\[
\delta \alpha_{i \,\rm out}(\omega)=\delta \alpha_i(\omega)
\frac{\gamma_i/2+i\,\omega}{\sqrt{\gamma_i}}-\frac{i\,g}{\sqrt{\gamma_i}}\delta
\Sigma_{i0}(\omega)\, .
\]
Using this last equation, the $\theta$-quadrature noise spectrum,
Eq.~(\ref{eq:quadraturenoise}), can be written as a linear combination
of the elements of the matrix $\boldsymbol{S}(\omega )$ given by
Eq.~(\ref{eq:spectrum2}). Since we are using c-numbers, the quadrature noise
spectrum calculated using this method is equivalent to the normal
ordered quadrature noise spectrum calculated with the original operator. 
The above results are only valid if the steady state solutions are stable. This can
be verified by calculating the eigenvalues of the matrix $\boldsymbol{B}$: If
they all are negative, then the system is stable.

\section{Results and analysis}
\label{sc:analysis}
\subsection{General Results}
We calculate here the $\theta$ quadrature noise spectrum of the output field,
given by Eq.~(\ref{eq:quadraturenoise}), when the incoming probe field is squeezed and the 
Rabi frequencies associated with each dipole
transition of the atom are different from zero.
We will suppose that $\Gamma_1=\Gamma_2=\Gamma$, $g_1=g_2=g$ and $\gamma_1=\gamma_2=\gamma$.
The initial conditions are as follows. Every mode of the pump field outside the
cavity is in vacuum state except the outside mode of the pump field which has the same
frequency as the mode $1$ inside the cavity. This last mode is in a coherent state such that the
mode one annihilation operator, $\hat a_1$, inside the cavity has mean value $\alpha_1$. This
means that $r_1=0$ in Eqs.~(\ref{eq:campoinicial}). Every mode of the probe field
is in squeezed vacuum for the $\theta_2=0$ quadrature, except the outside mode of the probe field with the same
frequency as mode two inside the cavity.  This last mode is in the squeezed state such that the
mode two annihilation operator, $\hat a_2$, inside the cavity has mean value
$\alpha_2$. This means that $r_2>0$ and $\theta_2=0$ in Eqs.~(\ref{eq:campoinicial}).

Under such initial conditions, Eq.~(\ref{eq:spectrum2}) can be calculated analytically with the help of
computer programs for symbolic algebra. Using the solution obtained with this
procedure and Eq.~(\ref{eq:quadraturenoise}), we calculate the
quadrature noise spectrum. The solution is
given in the appendix~\ref{ap:general} by
Eq.~(\ref{eq:yp}) and Eq.~(\ref{eq:yb}).

\begin{figure*}[!ht]
\subfigure[]{
\includegraphics[width=3in]{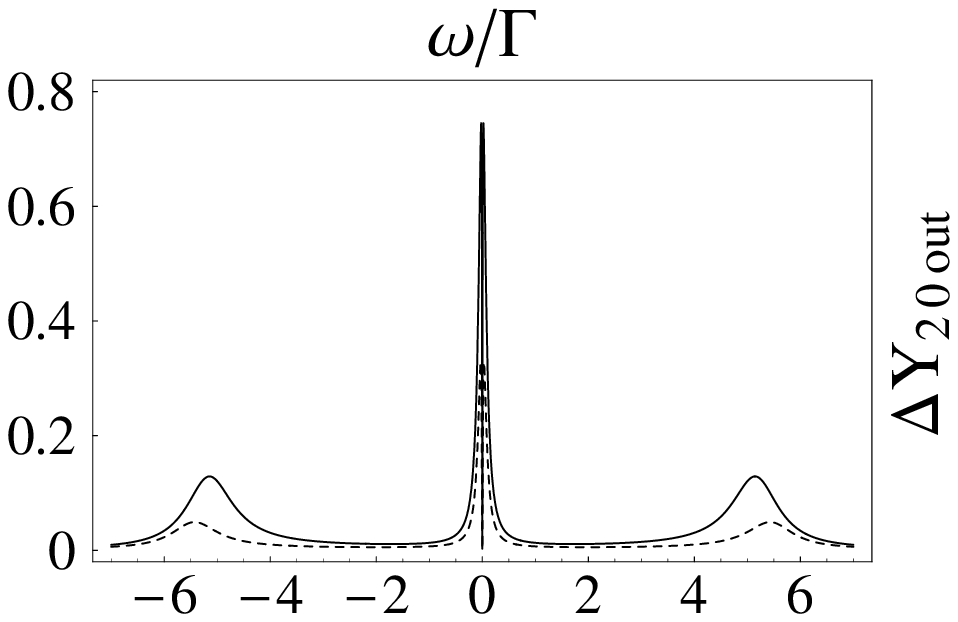}\label{fig:yipGa}}
\subfigure[]{\includegraphics[width=3in]{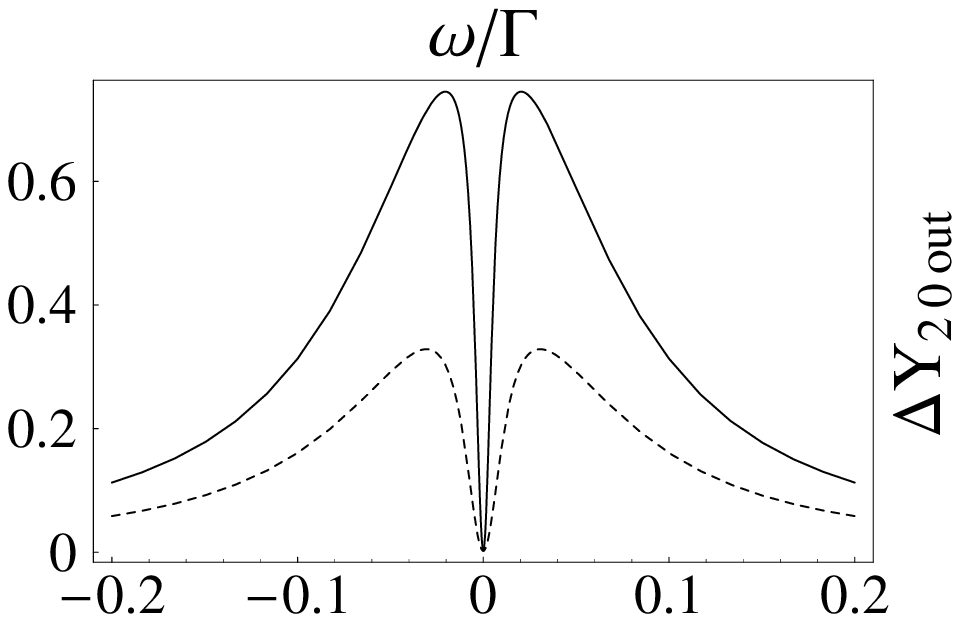}\label{fig:yipGb}}
\caption{\label{fig:yipG} Noise spectrum of $\theta=0$ quadrature of the
  output probe field. $C=167$, $\gamma=0.15 \Gamma$, $r_2=3$,
  $\Omega_1=\Omega_2=\Gamma$ (continuous line); $\Omega_1=\Gamma$, $\Omega_2=2\Gamma$ (dashed line).}
\end{figure*}
\begin{figure*}[!ht]
\subfigure[]{
\includegraphics[width=3in]{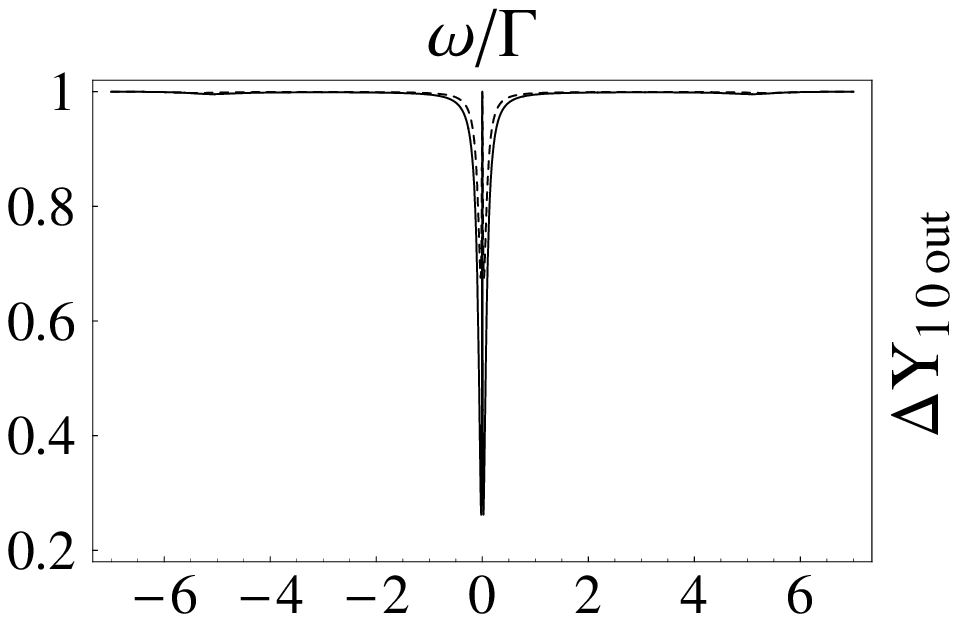}\label{fig:yiba}}
\subfigure[]{\includegraphics[width=3in]{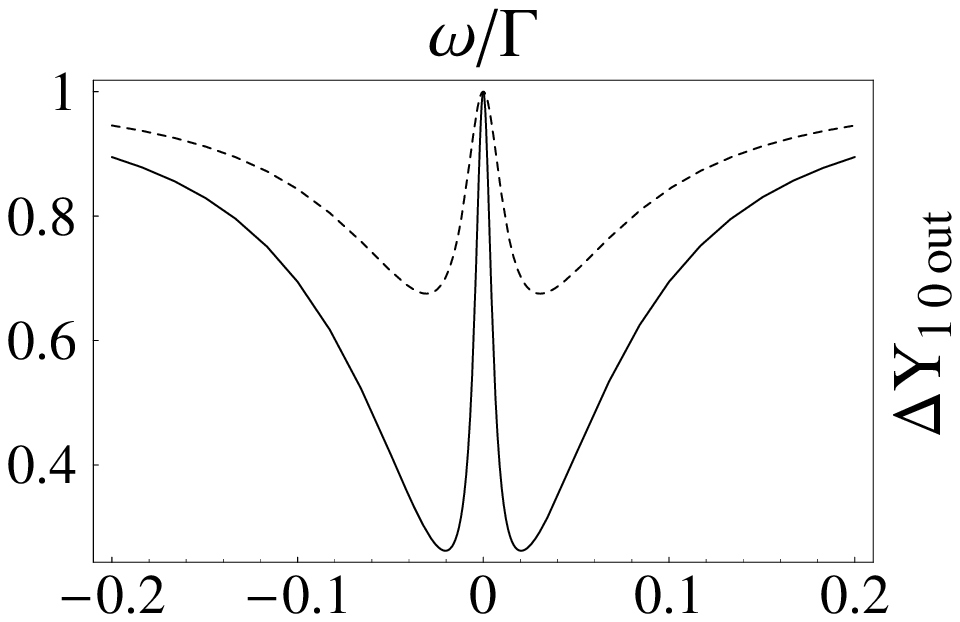}\label{fig:yibb}}
\caption{\label{fig:yib} Noise spectrum of $\theta=0$ quadrature of the
  output pump field. $C=167$, $\gamma=0.15 \Gamma$, $r_2=3$,
  $\Omega_1=\Omega_2=\Gamma$ (continuous line); $\Omega_1=\Gamma$,
  $\Omega_2=2\Gamma$ (dashed line). In Fig.~\ref{fig:yiba} a zoom of the behavior between
  $\omega/\Gamma=4$ and $6$ and between $\omega/\Gamma=-4$ and $-6$ is shown
  in the insets.}
\end{figure*}

\subsection{Analysis}
In order to gain some insight on what happens with the field after interacting with the atoms
with the given initial conditions, we start by plotting Eq.~(\ref{eq:yp}) and
Eq.~(\ref{eq:yb}). The bandwidth of the usual transparency curve in EIT
(defined outside the cavity) is a function of the Rabi frequencies. 
We will choose as the
Rabi frequencies of the plot, the interesting case in which the cavity bandwidth, characterized by $\gamma$, 
is smaller than the usual EIT bandwidth. 
This guarantees that the mean value of the modes of the field which enters the cavity is hardly altered by the atoms. Interesting
results appear in the case of large cooperation parameter $C=\frac{g^2 N}{\Gamma\gamma}\gg 1$.
In Fig.~\ref{fig:yipGa} we show a typical behavior of the $\theta=0$ quadrature noise
spectra of the probe field for the case of a good cavity ($\gamma=0.15\Gamma$), large
cooperation parameter $C=\frac{g^2 N}{\Gamma\gamma}=167\gg 1$, $r_2=3$, $\Omega_1=g_1\langle\alpha_1\rangle=\Gamma$,
$\Omega_2=g_2\langle\alpha_2\rangle=2\Gamma$ (dashed line) and
$\Omega_1=\Omega_2=\Gamma$ (continuous line).
 In Fig.~\ref{fig:yipGb} we show the behavior for small $\omega$. When
 $\omega\neq 0$ we observe, for each curve, four maxima where the initial
$\theta=0$ quadrature squeezing is absorbed. Two are located very close to
$\omega=0$. These maxima is located in a region where the absorption of the mean
value of the field is practically negligible ($\omega\ll\Omega_1,\Omega_2$). In the case
where $\Omega_1=\Omega_2$ (continuous line) the squeezing absorption is larger. 
The
other two maxima are for high ($|\omega|\gg\Omega_1,\Omega_2$) frequencies. For high frequencies we expect some absorption due to the vacuum
Rabi splitting. In a cavity filled with $N$ two level systems, the vacuum Rabi 
splitting is 
proportional to the square root of the cooperation parameter $C$
\cite{rv:agarwal2niveles} \cite{rv:raizen}. In the case we are studying we expect that the frequencies where 
these two maxima happens increases monotonically with $C$. This would be shown further on.

In Fig.~\ref{fig:yiba} and Fig.~\ref{fig:yibb} we show a typical behavior of the $\theta=0$ quadrature noise
spectra of the pump for the same parameters. In this case we have four
minimum at the same position of the four maxima in Fig.~\ref{fig:yipG}. We observe that an important amount of the
initial probe $\theta=0$ quadrature squeezing, absorbed by the medium in the two maxima close to $\omega=0$, goes
to the pump field for the same frequencies (compare Fig.~\ref{fig:yipGb} with
Fig.~\ref{fig:yibb}). In the case where $\Omega_1=\Omega_2$ (continuous line)
the squeezing transfer is larger. We also observe a transfer of
squeezing from probe to pump field in the two maxima far from the origin of
Fig.~\ref{fig:yipGa}, but in this case it is rather small. We point out the two
major characteristics of the behavior of the quadrature spectra showed in Fig.~\ref{fig:yipG} and Fig.~\ref{fig:yib}: a) the
two maxima, locate near $\omega=0$, of $\theta=0$ squeezing absorption is for frequencies where the
mean value of the field is hardly altered; and b) we have a transfer of
squeezing between the probe and pump field. To the best of our knowledge this is the first
theoretical work that describes such squeezing transfer.  
In the rest of this section we will characterize the behavior of these maxima
and minima as a function of the parameters of the system.

We now obtain expressions for the position and values of these maxima in order
to understand their behavior as a function of the system parameters. 
The first maximum can be obtained using
Eq.~(\ref{eq:yp}). If we suppose that 
$(\Omega_1^2+\Omega_2^2) \gamma^2\ll (C\gamma\Gamma+\Omega_1^2+\Omega_2^2)^2$, $C\gg 1$,
$\gamma/\Gamma\ll 1$, then the frequency where this maximum occurs can be 
approximated as
\begin{equation}\label{eq:maxima1}
\omega_{max}\approx \pm\frac{\gamma\sqrt{\Omega _1^2+\Omega _2^2}}{2\sqrt{C
    \gamma +\Omega _1^2+\Omega _2^2}}\, ,
\end{equation} 
with the $\theta$ quadrature noise for this frequency given by
\begin{equation}\label{eq:yipmax}
\Delta Y_{2\,\theta\,\rm {out}}(\omega_{max})\approx f(\theta)+\frac{4 C^2\gamma ^2\Gamma^2 \Omega _1^2 \Omega _2^2 \left(1-f(\theta)\right) }{\left(\Omega _1^2+\Omega _2^2\right)^2 \left(2 \Omega _1^2+2 \Omega _2^2+C \gamma\Gamma
   \right)^2}\, ,
\end{equation}
and
\begin{equation}\label{eq:yibmax}
\Delta Y_{1\,\theta\,\rm out}(\omega_{max})\approx 1-\frac{4 C^2\gamma ^2\Gamma^2 \Omega _1^2 \Omega _2^2 \left(1-f(\theta)\right) }{\left(\Omega _1^2+\Omega _2^2\right)^2 \left(2 \Omega _1^2+2 \Omega _2^2+C \gamma\Gamma
   \right)^2}\, ,
\end{equation}
where
\[
f(\theta)=e^{-2\,r_2}\cos^2\theta+e^{2\,r_2}\sin^2\theta\, .
\]
From Eq.~(\ref{eq:maxima1}),  we have that 
$\omega_{max}<\gamma/2$.  In the case where $\gamma$ is smaller than the Rabi
frequencies associated with the atoms optical transitions, the mean value of
the field which enters the cavity is almost unaltered. Nevertheless, because $\omega_{max}<\gamma/2$, we can conclude that, at least for an initially 
squeezed state, there is huge alteration of the initial quantum fluctuations
of the field after interacting with the medium.

Comparing Eq.~(\ref{eq:yipmax}) with Eq.~(\ref{eq:yibmax}) when $\theta=0$ and
$\Omega_1,\Omega_2\neq 0$ it can be seen that there is 
a transfer of quadrature squeezing, for that particular frequency, from the probe to
the pump field. This transfer of squeezing is maximum in the
case where $\Omega_1=\Omega_2=\Omega$. For this case, and when
$C\gamma\gg\Omega$, Eq.~(\ref{eq:yipmax}) gives $\Delta Y_{2\,0\,\rm
  {out}}(\omega_{max})\approx 1$ and Eq.~(\ref{eq:yibmax}) gives $\Delta
Y_{1\,0\,\rm out}(\omega_{max})\approx e^{-2 r_2}$. In this last case the effect of the
interaction with the atoms is to transfer all the
initial squeezing from the probe field to the pump field for the
frequency $\omega_{max}$. For other
quadratures and the same conditions for the parameters, we obtain  from Eq.~(\ref{eq:yipmax}) that
$\Delta Y_{2\,0\,\rm
  {out}}(\omega_{max})\approx 1$ and from Eq.~(\ref{eq:yibmax}) that $\Delta
Y_{1\,0\,\rm out}(\omega_{max})\approx f(\theta)$. This means that the initial
fluctuations for each quadrature of the probe field have been transferred to the pump field. When $\Omega_2$  is zero
there is no such transfer of fluctuations and the medium is transparent for the initial
squeezed vacuum. This transparency for a squeezed vacuum probe field in cavity
EIT was shown theoretically in \cite{rv:dantan3} \cite{rv:dantan4}.

The $\theta=0$ quadrature noise spectrum, Eq.~(\ref{eq:yp}), has two maxima,
which corresponds to the two maxima for $|\omega|>\Omega_1,\Omega_2$ in Fig.~\ref{fig:yipG}. When $\Gamma\ll
\Omega_1$ or $\Gamma\ll
\Omega_2$ or $\Gamma\ll C \gamma$  the positions of these two maxima can be approximated by
 \[
\omega'_{max}\approx\pm\sqrt{\Omega _1^2+\Omega _2^2+C \gamma \Gamma}\, .
\]
As can be seen from the last equation, $|\omega'_{max}|$ increases
monotonically with the cooperation parameter. These maxima
correspond to the maxima we expected due to the 
vacuum Rabi splitting and mentioned before.
The quadrature noise spectrum for these maxima position is
\begin{equation}\label{eq:yipmax2}
\Delta Y_{2\,\theta\,\rm out}(\omega'_{max})\approx f(\theta)+\frac{2 C \left(1-f(\theta)\right) \gamma^2 \Omega _1^2}{\left(\Omega _1^2+\Omega _2^2\right) \left(\Omega _1^2+\Omega _2^2+C \gamma
   \Gamma\right)}\, ,
\end{equation}

\begin{equation}\label{eq:yibmax2}
\Delta Y_{1\,\theta\,\rm out}(\omega'_{max})\approx 1-\frac{C^2 \left(1-f(\theta)\right) \gamma ^4 \Omega _1^2 \Omega _2^2}{\left(\Omega _1^2+\Omega _2^2\right)^2 \left(\Omega _1^2+\Omega _2^2+C \gamma 
   \Gamma\right)^2} \, .
\end{equation}
From Eq.~(\ref{eq:yipmax2}) and Eq.~(\ref{eq:yibmax2}), we may conclude that
transfer of squeezing from the probe to the pump occurs for these
frequencies, but it is much smaller, and can not be perfect, contrary to the
situation corresponding to 
the two maxima close to $\omega=0$ (compare
Fig.~\ref{fig:yipGa} with Fig.~\ref{fig:yiba}).  

Using the DGCZ inequality for continuum variables \cite{rv:dgcz}, we looked
for conditions under which we could guarantee  the existence of
quantum correlations between the pump and probe quadratures. Using
Eqs.~(\ref{eq:yp})~(\ref{eq:yb}) for the quadratures noise spectrum and Eq.~(\ref{eq:co})
for the correlations noise spectrum, we looked for parameters under which the
inequality could be violated. For spectrum frequencies $\omega_{max}$ and $\omega'_{max}$ it is not difficult to see that 
the inequality is never violated. We numerically looked for the inequality violation 
for parameters $C$ between $10$ and $200$, $\Omega_1$ and $\Omega_2$ between
$\Gamma$ and $10\Gamma$, $\gamma=0.15\Gamma$ and spectrum frequencies $\omega$
between $-10\Gamma$ and $10\Gamma$.
We did not find any quantum correlations between the pump and
probe field. Nevertheless, the fact that for some spectrum frequencies both fields are
squeezed after interaction, is a signature of quantum correlations between
 some orthogonal modes (not necesarily the pump and probe modes). A method
to find the modes possessing EPR-type correlations is given in
\cite{rv:dantan1}.

\section{Conclusions}
\label{sc:conclusions}
There has been much recent activity, both theoretical and experimental, to study
the properties of quantum states interacting with three-level atoms presenting
EIT. The aim of this activity has been to coherently control the propagation of quantum
light. Interesting applications arise, as for example quantum information
processing \cite{rv:rmpfleisch}. In this paper we studied the output field quadrature noise spectra
of an initially squeezed probe field after interacting with three-level atoms
presenting EIT.   
In cavity EIT, there is high alteration of the initial quantum properties of the probe field even when the
mean values of the field are unaltered. This is shown with initially squeezed states for
the probe field. We showed that there exists two frequencies where the maximum
absorption of quantum fluctuation occurs, that can be very close to atom resonance,
even when the Rabi frequencies are much larger than the other
parameters. Moreover, some of the squeezing absorbed in the probe field can be transferred
to the pump field. This transfer is maximum when the Rabi frequencies
associated with each field are equal. It is almost perfect if in addition $C\gg
1$.  This transfer implies a coherent exchange of quantum
properties between the probe and pump field. Using the DGCZ inequality we did not find any quantum correlations between the pump and probe
field after interacting with the atoms. We expect that these results may be useful to
predict and explain experimental results where the quadrature noise spectrum is
measured after interacting with atoms presenting EIT.

\acknowledgments
We thank M. Bienert for revising the manuscript and fruitful discussions. This
work was supported by the Mexican agency CONACYT under the project 41000-F.

\appendix
\section{Diffusion coefficients}
\label{sc:difusion}
The non-zero diffusion coefficients, obtained using the generalized Einstein
relations, are given by
\begin{eqnarray}
D_{W_1W_1}&=&(4\Gamma_1+\Gamma_2)\langle\hat\Sigma_{00}\rangle,\nonumber\\
D_{W_2W_2}&=&(\Gamma_1+4\Gamma_2)\langle\hat\Sigma_{00}\rangle,\nonumber\\
D_{W_1W_2}&=&D_{W_2W_1}=(2\Gamma_1+2\Gamma_2)\langle\hat\Sigma_{00}\rangle,\nonumber\\
D_{W_1\Sigma_{01}}&=&D_{\Sigma_{10}W_1}^*=(2\Gamma_1+\Gamma_2)\langle\hat\Sigma_{01}\rangle,\nonumber\\
D_{W_1\Sigma_{02}}&=&D_{\Sigma_{20}W_1}^*=(2\Gamma_1+\Gamma_2)\langle\hat\Sigma_{02}\rangle,\nonumber\\
D_{W_2\Sigma_{01}}&=&D_{\Sigma_{10}W_2}^*=(\Gamma_1+2\Gamma_2)\langle\hat\Sigma_{01}\rangle,\nonumber\\
D_{W_2\Sigma_{02}}&=&D_{\Sigma_{20}W_2}^*=(\Gamma_1+2\Gamma_2)\langle\hat\Sigma_{02}\rangle,\nonumber\\
D_{\Sigma_{10}\Sigma_{01}}&=&\Gamma_1\langle \hat\Sigma_{00}\rangle+(\Gamma_1+\Gamma_2)\langle \hat\Sigma_{11}\rangle, \nonumber\\
D_{\Sigma_{20}\Sigma_{02}}&=&\Gamma_2\langle \hat\Sigma_{00}\rangle+(\Gamma_1+\Gamma_2)\langle \hat\Sigma_{22}\rangle, \nonumber\\
D_{\Sigma_{01}\Sigma_{12}}&=&D_{\Sigma_{21}\Sigma_{10}}^*=\Gamma_{12}\, \langle\hat\Sigma_{02}\rangle,\nonumber\\
D_{\Sigma_{02}\Sigma_{21}}&=&D_{\Sigma_{12}\Sigma_{20}}^*=\Gamma_{12}\, \langle\hat\Sigma_{01}\rangle,\nonumber\\
D_{\Sigma_{12}\Sigma_{21}}&=&\Gamma_1\langle\hat\Sigma_{00}\rangle+2\Gamma_{12}\langle\hat\Sigma_{11}\rangle, \nonumber\\
D_{\Sigma_{21}\Sigma_{12}}&=&\Gamma_1\langle\hat\Sigma_{00}\rangle+2\Gamma_{12}\langle\hat\Sigma_{22}\rangle,
\nonumber\\ 
D_{\Sigma_{10}\Sigma_{02}}&=&D_{\Sigma_{20}\Sigma_{01}}^*=(\Gamma_1+\Gamma_2-\Gamma_{12})\langle \hat\Sigma_{12}\rangle. 
\label{eq:difcoecav}
\end{eqnarray}

The non-zero c-number diffusion coefficients, obtained by transforming the
operator Eqs~(\ref{eq:sistemacav}) into an equivalent set of c-number equations, are given by
\begin{eqnarray}
D_{W_1W_1}&=&(4\Gamma_1+\Gamma_2) \langle\Sigma_{00}\rangle\nonumber \\&&-i(4
\Omega_1\langle\Sigma_{01}\rangle+\Omega_2\langle\Sigma_{02}\rangle-{\rm c.c.}),\nonumber\\
D_{W_2W_2}&=&(\Gamma_1+4\Gamma_2)\langle\Sigma_{00}\rangle\nonumber \\&&-i(4 \Omega_2\langle\Sigma_{02}\rangle+\Omega_1\langle\Sigma_{01}\rangle-{\rm c.c.}),\nonumber\\
D_{W_1W_2}&=&(2\Gamma_1+2\Gamma_2)\langle\Sigma_{00}\rangle\nonumber \\&&-2 i(\Omega_2\langle\Sigma_{02}\rangle+\Omega_1\langle\Sigma_{01}\rangle-{\rm c.c.}),\nonumber\\
D_{\Sigma_{12}\Sigma_{21}}&=&\Gamma_1\langle\Sigma_{00}\rangle+2\Gamma_{12}\langle\Sigma_{11}\rangle-(i\Omega_2\langle\Sigma_{12}\rangle+{\rm c.c.}),\nonumber\\
D_{\Sigma_{02}\Sigma_{12}}&=&D_{\Sigma_{20}\Sigma_{21}}^*=- i\Omega_2^*\langle\Sigma_{12}\rangle,\nonumber\\
D_{\Sigma_{02}\Sigma_{21}}&=&D_{\Sigma_{20}\Sigma_{12}}^*=\Gamma_{12}\, \langle\Sigma_{01}\rangle,\nonumber\\
D_{W_1\Sigma_{10}}&=&D_{W_1\Sigma_{01}}^*=i\Omega_2\langle\Sigma_{12}\rangle,\nonumber\\
D_{W_1\Sigma_{21}}&=&D_{W_1\Sigma_{12}}^*=-2 i\Omega_1^*\langle\Sigma_{20}\rangle+2
i\Omega_2\langle\Sigma_{01}\rangle,\nonumber\\
D_{W_2\Sigma_{10}}&=&D_{W_2\Sigma_{01}}^*=-i\Omega_2\langle\Sigma_{12}\rangle,\nonumber\\
D_{W_2\Sigma_{21}}&=&D_{W_2\Sigma_{12}}^*=-i\Omega_1^*\langle\Sigma_{20}\rangle+i\Omega_2\langle\Sigma_{01}\rangle,\nonumber\\
D_{\Sigma_{10}\Sigma_{10}}&=&D_{\Sigma_{01}\Sigma_{01}}^*=2 i\Omega_1\langle\Sigma_{10}\rangle,\nonumber\\
D_{\Sigma_{01}\Sigma_{02}}&=&D_{\Sigma_{20}\Sigma_{10}}^*=-i\Omega_1^*\langle\Sigma_{02}\rangle-i\Omega_2^*\langle\Sigma_{01}\rangle,\nonumber\\
D_{\Sigma_{20}\Sigma_{20}}&=&D_{\Sigma_{02}\Sigma_{02}}^*=2 i\Omega_2\langle\Sigma_{20}\rangle,\nonumber\\
D_{\Sigma_{01}\Sigma_{12}}&=&D_{\Sigma_{21}\Sigma_{10}}^*\nonumber \\
&=&{\Gamma_{12}\, \langle\Sigma_{02}\rangle+i\Omega_2(\langle
  W_1\rangle-\langle W_2\rangle)+i\Omega_1\langle\Sigma_{12}\rangle}\, ,\nonumber \\
\label{eq:cnumbercoe}
\end{eqnarray}
where $\Omega_i=g_i\,\langle\alpha_i\rangle$.
\section{General results}
\label{ap:general}
The general results for the $\theta$ quadrature noise spectrum, defined in
Eq.~(\ref{eq:quadraturenoise}), are 
\begin{widetext}
\begin{eqnarray}\label{eq:yp}
&& \Delta Y_{2\,\theta\,\rm{out}}(\omega')=\Big(8 C \Gamma ^2 \omega ^2 \gamma ^2 \left( 2 C  \Omega _1^2 \Omega _2^2 \gamma ^2+ \left(\gamma ^2+4 \omega ^2\right) \Omega _1^2 \left(\Omega _1^2+\Omega
   _2^2\right) \right) \nonumber\\&&+f(\theta) \left(-4 C \gamma^2  \Gamma^2 \omega ^2
   \left(\gamma ^2+4 \omega ^2\right) \left(\Omega _1^2+\Omega _2^2\right)
   \left(\Omega _1^2-\Omega _2^2\right)+A+B\right)\Big)/M\, ,\nonumber\\
\end{eqnarray}
\begin{eqnarray}\label{eq:yb}
&& \Delta Y_{1\,\theta\,\rm{out}}(\omega')=\Big(16 C^2  \Gamma ^2 \omega ^2
\Omega _1^2 \Omega _2^2 \gamma ^4\,f(\theta)+A+B\Big)/M\, ,
\end{eqnarray}
where
\begin{equation}
A=4 C^2 \Gamma ^2 \omega ^2  \gamma ^2 \left(\gamma ^2 \left(\Omega
   _1^2-\Omega _2^2\right)^2+4 \omega ^2 \left(\Omega _1^2+\Omega
   _2^2\right)^2\right)\nonumber\, ,
\end{equation}
\begin{eqnarray}
&&B=\left(\gamma
   ^2+4 \omega ^2\right) \left(\Omega _1^2+\Omega _2^2\right)^2 \Big\{4 C \Gamma  \omega ^2 \gamma  \left(-2 \omega ^2+2 \Omega _1^2+2 \Omega _2^2+\gamma  \Gamma \right) \nonumber\\&& +\left(\gamma ^2+4
   \omega ^2\right) \left(\omega ^2 \left(\Gamma ^2+\omega ^2\right)-\left(2 \omega ^2-\Omega _1^2-\Omega _2^2\right)
   \left(\Omega _1^2+\Omega _2^2\right)\right)\Big\}\nonumber\, ,
\end{eqnarray}
\begin{equation}
M=4 C^2 \gamma ^2 \Gamma ^2 \omega ^2 \left(\gamma ^2+4 \omega ^2\right)
\left(\Omega _1^2+\Omega _2^2\right)^2+B\nonumber\, ,
\end{equation}
\begin{equation}
f(\theta)=e^{-2\,r_2}\cos^2\theta+e^{2\,r_2}\sin^2\theta\nonumber\, .
\end{equation}
The correlation between the $\theta_1$ quadrature noise spectrum of the probe
field and the $\theta_2$ quadrature noise spectrum of the pump
field is
\begin{eqnarray}\label{eq:co}
&& \Delta C_{\theta_1,\theta_2\,
   \rm{out}}(\omega')=\langle\delta Y_{1\,\theta_1 \,\rm out}(\omega)  \delta Y_{2\,\theta_2 \,\rm out}(\omega')\rangle=\nonumber\\&&f_2(\theta_1,\theta_2) 8 \Gamma ^2 \gamma ^2 \omega ^2
   \Omega _1 \Omega _2  C \left(\left(\gamma ^2+4
   \omega ^2\right)  \left(\Omega _1^2+\Omega _2^2\right)-2 C \gamma ^2
   \left(\Omega _1^2-\Omega _2^2\right)\right)/M\, ,
\end{eqnarray}
where
\begin{equation}
f_2(\theta_1,\theta_2)=(e^{-2 r_2} \cos \theta_1 \cos \theta_2+e^{2 r_2} \sin \theta_1 \sin
   \theta_2-\cos (\theta_1-\theta_2))/2\, .
\end{equation}
\end{widetext}


\begin{thebibliography}{23}
\expandafter\ifx\csname natexlab\endcsname\relax\def\natexlab#1{#1}\fi
\expandafter\ifx\csname bibnamefont\endcsname\relax
  \def\bibnamefont#1{#1}\fi
\expandafter\ifx\csname bibfnamefont\endcsname\relax
  \def\bibfnamefont#1{#1}\fi
\expandafter\ifx\csname citenamefont\endcsname\relax
  \def\citenamefont#1{#1}\fi
\expandafter\ifx\csname url\endcsname\relax
  \def\url#1{\texttt{#1}}\fi
\expandafter\ifx\csname urlprefix\endcsname\relax\def\urlprefix{URL }\fi
\providecommand{\bibinfo}[2]{#2}
\providecommand{\eprint}[2][]{\url{#2}}

\bibitem[{\citenamefont{Harris}(1997)}]{rv:harris}
\bibinfo{author}{\bibfnamefont{S.}~\bibnamefont{Harris}},
  \bibinfo{journal}{Phys. Today} \textbf{\bibinfo{volume}{50}},
  \bibinfo{pages}{36} (\bibinfo{year}{1997}).

\bibitem[{\citenamefont{Marangos}(1998)}]{rv:marangos}
\bibinfo{author}{\bibfnamefont{J.~P.} \bibnamefont{Marangos}},
  \bibinfo{journal}{J. Mod. Opt.} \textbf{\bibinfo{volume}{45}},
  \bibinfo{pages}{471} (\bibinfo{year}{1998}).

\bibitem[{\citenamefont{Hau et~al.}(1999)\citenamefont{Hau, Harris, Dutton, and
  Behroozi}}]{rv:hau}
\bibinfo{author}{\bibfnamefont{L.~V.} \bibnamefont{Hau}},
  \bibinfo{author}{\bibfnamefont{S.~E.} \bibnamefont{Harris}},
  \bibinfo{author}{\bibfnamefont{Z.}~\bibnamefont{Dutton}}, \bibnamefont{and}
  \bibinfo{author}{\bibfnamefont{C.~H.} \bibnamefont{Behroozi}},
  \bibinfo{journal}{Nature} \textbf{\bibinfo{volume}{397}},
  \bibinfo{pages}{594} (\bibinfo{year}{1999}).

\bibitem[{\citenamefont{Liu et~al.}(2001)\citenamefont{Liu, Dutton, Behroozi,
  and Hau}}]{rv:liu}
\bibinfo{author}{\bibfnamefont{C.}~\bibnamefont{Liu}},
  \bibinfo{author}{\bibfnamefont{Z.}~\bibnamefont{Dutton}},
  \bibinfo{author}{\bibfnamefont{C.~H.} \bibnamefont{Behroozi}},
  \bibnamefont{and} \bibinfo{author}{\bibfnamefont{L.~V.} \bibnamefont{Hau}},
  \bibinfo{journal}{Nature} \textbf{\bibinfo{volume}{409}},
  \bibinfo{pages}{490} (\bibinfo{year}{2001}).

\bibitem[{\citenamefont{Fleischhauer and Lukin}(2002)}]{rv:memoria2}
\bibinfo{author}{\bibfnamefont{M.}~\bibnamefont{Fleischhauer}}
  \bibnamefont{and} \bibinfo{author}{\bibfnamefont{M.~D.} \bibnamefont{Lukin}},
  \bibinfo{journal}{Phys. Rev. A} \textbf{\bibinfo{volume}{65}},
  \bibinfo{pages}{022314} (\bibinfo{year}{2002}).

\bibitem[{\citenamefont{Akamatsu et~al.}(2004)\citenamefont{Akamatsu, Akiba,
  and Kozuma}}]{rv:kozuma}
\bibinfo{author}{\bibfnamefont{D.}~\bibnamefont{Akamatsu}},
  \bibinfo{author}{\bibfnamefont{K.}~\bibnamefont{Akiba}}, \bibnamefont{and}
  \bibinfo{author}{\bibfnamefont{M.}~\bibnamefont{Kozuma}},
  \bibinfo{journal}{Phys. Rev. Lett.} \textbf{\bibinfo{volume}{92}},
  \bibinfo{pages}{203602} (\bibinfo{year}{2004}).

\bibitem[{\citenamefont{Alzar et~al.}(2003)\citenamefont{Alzar, Cruz, G\'omez,
  Santos, and Nussenzveig}}]{rv:carlosruido}
\bibinfo{author}{\bibfnamefont{C.~L.~G.} \bibnamefont{Alzar}},
  \bibinfo{author}{\bibfnamefont{L.~S.} \bibnamefont{Cruz}},
  \bibinfo{author}{\bibfnamefont{J.~G.~A.} \bibnamefont{G\'omez}},
  \bibinfo{author}{\bibfnamefont{M.~F.} \bibnamefont{Santos}},
  \bibnamefont{and}
  \bibinfo{author}{\bibfnamefont{P.}~\bibnamefont{Nussenzveig}},
  \bibinfo{journal}{Europhys. Lett.} \textbf{\bibinfo{volume}{61}},
  \bibinfo{pages}{485} (\bibinfo{year}{2003}).

\bibitem[{\citenamefont{Martinelli et~al.}(2004)\citenamefont{Martinelli,
  Valente, Failache, Felinto, Cruz, Nussenzveig, and Lezama}}]{rv:lezama}
\bibinfo{author}{\bibfnamefont{M.}~\bibnamefont{Martinelli}},
  \bibinfo{author}{\bibfnamefont{P.}~\bibnamefont{Valente}},
  \bibinfo{author}{\bibfnamefont{H.}~\bibnamefont{Failache}},
  \bibinfo{author}{\bibfnamefont{D.}~\bibnamefont{Felinto}},
  \bibinfo{author}{\bibfnamefont{L.~S.} \bibnamefont{Cruz}},
  \bibinfo{author}{\bibfnamefont{P.}~\bibnamefont{Nussenzveig}},
  \bibnamefont{and} \bibinfo{author}{\bibfnamefont{A.}~\bibnamefont{Lezama}},
  \bibinfo{journal}{Phys. Rev. A} \textbf{\bibinfo{volume}{69}},
  \bibinfo{pages}{043809} (\bibinfo{year}{2004}).

\bibitem[{\citenamefont{Sautenkov et~al.}(2005)\citenamefont{Sautenkov,
  Rostovtsev, and Scully}}]{rv:scullyeitco}
\bibinfo{author}{\bibfnamefont{V.~A.} \bibnamefont{Sautenkov}},
  \bibinfo{author}{\bibfnamefont{Y.~V.} \bibnamefont{Rostovtsev}},
  \bibnamefont{and} \bibinfo{author}{\bibfnamefont{M.~O.}
  \bibnamefont{Scully}}, \bibinfo{journal}{Phys. Rev. A}
  \textbf{\bibinfo{volume}{72}}, \bibinfo{pages}{065801}
  (\bibinfo{year}{2005}).

\bibitem[{\citenamefont{Barberis-Blostein and Zagury}(2004)}]{rv:pablonicim}
\bibinfo{author}{\bibfnamefont{P.}~\bibnamefont{Barberis-Blostein}}
  \bibnamefont{and} \bibinfo{author}{\bibfnamefont{N.}~\bibnamefont{Zagury}},
  \bibinfo{journal}{Phys. Rev. A} \textbf{\bibinfo{volume}{70}},
  \bibinfo{pages}{053827} (\bibinfo{year}{2004}).

\bibitem[{\citenamefont{Dantan and Pinard}(2004)}]{rv:dantan3}
\bibinfo{author}{\bibfnamefont{A.}~\bibnamefont{Dantan}} \bibnamefont{and}
  \bibinfo{author}{\bibfnamefont{M.}~\bibnamefont{Pinard}},
  \bibinfo{journal}{Phys. Rev. A} \textbf{\bibinfo{volume}{69}},
  \bibinfo{pages}{043810} (\bibinfo{year}{2004}).

\bibitem[{\citenamefont{Dantan et~al.}(2005)\citenamefont{Dantan, Bramati, and
  Pinard}}]{rv:dantan4}
\bibinfo{author}{\bibfnamefont{A.}~\bibnamefont{Dantan}},
  \bibinfo{author}{\bibfnamefont{A.}~\bibnamefont{Bramati}}, \bibnamefont{and}
  \bibinfo{author}{\bibfnamefont{M.}~\bibnamefont{Pinard}},
  \bibinfo{journal}{Phys. Rev. A} \textbf{\bibinfo{volume}{71}},
  \bibinfo{pages}{043801} (\bibinfo{year}{2005}).

\bibitem[{\citenamefont{Dantan et~al.}()\citenamefont{Dantan, Cviklinski,
  Giacobino, and Pinard}}]{rv:dantan5}
\bibinfo{author}{\bibfnamefont{A.}~\bibnamefont{Dantan}},
  \bibinfo{author}{\bibfnamefont{J.}~\bibnamefont{Cviklinski}},
  \bibinfo{author}{\bibfnamefont{E.}~\bibnamefont{Giacobino}},
  \bibnamefont{and} \bibinfo{author}{\bibfnamefont{M.}~\bibnamefont{Pinard}},
  \eprint{quant-ph/0603197}.

\bibitem[{\citenamefont{Walls and Milburn}(1994)}]{lb:wallsmilburn}
\bibinfo{author}{\bibfnamefont{D.~F.} \bibnamefont{Walls}} \bibnamefont{and}
  \bibinfo{author}{\bibfnamefont{G.~J.} \bibnamefont{Milburn}},
  \emph{\bibinfo{title}{Quantum Optics}} (\bibinfo{publisher}{Springer,
  Berlin}, \bibinfo{year}{1994}).

\bibitem[{\citenamefont{Orszag}(2000)}]{lb:orszag}
\bibinfo{author}{\bibfnamefont{M.}~\bibnamefont{Orszag}},
  \emph{\bibinfo{title}{Quantum Optics}} (\bibinfo{publisher}{Springer-Verlag},
  \bibinfo{year}{2000}).

\bibitem[{\citenamefont{Cohen-Tannoudji
  et~al.}(1992)\citenamefont{Cohen-Tannoudji, Dupont-Roc, and
  Grynberg}}]{lb:cohen}
\bibinfo{author}{\bibfnamefont{C.}~\bibnamefont{Cohen-Tannoudji}},
  \bibinfo{author}{\bibfnamefont{J.}~\bibnamefont{Dupont-Roc}},
  \bibnamefont{and} \bibinfo{author}{\bibfnamefont{G.}~\bibnamefont{Grynberg}},
  \emph{\bibinfo{title}{Atom-Photon Interactions}} (\bibinfo{publisher}{John
  Wiley \& Sons}, \bibinfo{year}{1992}).

\bibitem[{\citenamefont{Davidovich}(1996)}]{rv:davidovich}
\bibinfo{author}{\bibfnamefont{L.}~\bibnamefont{Davidovich}},
  \bibinfo{journal}{Rev. Mod. Phys.} \textbf{\bibinfo{volume}{68}},
  \bibinfo{pages}{127} (\bibinfo{year}{1996}).

\bibitem[{\citenamefont{Gardiner}(1994)}]{lb:gardiner}
\bibinfo{author}{\bibfnamefont{C.~W.} \bibnamefont{Gardiner}},
  \emph{\bibinfo{title}{Handbook of Stochastic Methods}}
  (\bibinfo{publisher}{Springer-Verlag}, \bibinfo{year}{1994}).

\bibitem[{\citenamefont{Agarwal}(1984)}]{rv:agarwal2niveles}
\bibinfo{author}{\bibfnamefont{G.~S.} \bibnamefont{Agarwal}},
  \bibinfo{journal}{Phys. Rev. Lett.} \textbf{\bibinfo{volume}{53}},
  \bibinfo{pages}{1732} (\bibinfo{year}{1984}).

\bibitem[{\citenamefont{Raizen et~al.}(1989)\citenamefont{Raizen, Thompson,
  Brecha, Kimble, and Carmichael}}]{rv:raizen}
\bibinfo{author}{\bibfnamefont{M.~G.} \bibnamefont{Raizen}},
  \bibinfo{author}{\bibfnamefont{R.~J.} \bibnamefont{Thompson}},
  \bibinfo{author}{\bibfnamefont{R.~J.} \bibnamefont{Brecha}},
  \bibinfo{author}{\bibfnamefont{H.~J.} \bibnamefont{Kimble}},
  \bibnamefont{and} \bibinfo{author}{\bibfnamefont{H.~J.}
  \bibnamefont{Carmichael}}, \bibinfo{journal}{Phys. Rev. Lett.}
  \textbf{\bibinfo{volume}{63}}, \bibinfo{pages}{240} (\bibinfo{year}{1989}).

\bibitem[{\citenamefont{Duan et~al.}(2000)\citenamefont{Duan, Giedke, Cirac,
  and Zoller}}]{rv:dgcz}
\bibinfo{author}{\bibfnamefont{L.-M.} \bibnamefont{Duan}},
  \bibinfo{author}{\bibfnamefont{G.}~\bibnamefont{Giedke}},
  \bibinfo{author}{\bibfnamefont{J.~I.} \bibnamefont{Cirac}}, \bibnamefont{and}
  \bibinfo{author}{\bibfnamefont{P.}~\bibnamefont{Zoller}},
  \bibinfo{journal}{Phys. Rev. Lett.} \textbf{\bibinfo{volume}{84}},
  \bibinfo{pages}{2722} (\bibinfo{year}{2000}).

\bibitem[{\citenamefont{Josse et~al.}(2004)\citenamefont{Josse, Dantan,
  Bramati, and Giacobino}}]{rv:dantan1}
\bibinfo{author}{\bibfnamefont{V.}~\bibnamefont{Josse}},
  \bibinfo{author}{\bibfnamefont{A.}~\bibnamefont{Dantan}},
  \bibinfo{author}{\bibfnamefont{A.}~\bibnamefont{Bramati}}, \bibnamefont{and}
  \bibinfo{author}{\bibfnamefont{E.}~\bibnamefont{Giacobino}},
  \bibinfo{journal}{J. Opt. B: Quantum Semiclass. Opt.}
  \textbf{\bibinfo{volume}{6}}, \bibinfo{pages}{532} (\bibinfo{year}{2004}).

\bibitem[{\citenamefont{Fleischhauer et~al.}(2005)\citenamefont{Fleischhauer,
  Imamoglu, and Marangos}}]{rv:rmpfleisch}
\bibinfo{author}{\bibfnamefont{M.}~\bibnamefont{Fleischhauer}},
  \bibinfo{author}{\bibfnamefont{A.}~\bibnamefont{Imamoglu}}, \bibnamefont{and}
  \bibinfo{author}{\bibfnamefont{J.~P.} \bibnamefont{Marangos}},
  \bibinfo{journal}{Rev. Mod. Phys.} \textbf{\bibinfo{volume}{77}},
  \bibinfo{pages}{633} (\bibinfo{year}{2005}).

\end{thebibliography}
\end{document}